\documentclass{svproc}
\usepackage{url}
\usepackage{graphicx}
\usepackage[lined,ruled,commentsnumbered]{algorithm2e}
\usepackage{amsmath}
\DeclareMathOperator*{\argmax}{arg\,max}

\begin{document}
\mainmatter              
\title{An Adaptive Hybrid Active Learning Strategy with Free Ratings in Collaborative Filtering}
\titlerunning{An Adaptive Hybrid Active Learning Strategy with Free Ratings in CF}  

\author{Alireza Gharahighehi\inst{1,2} \and Felipe Kenji Nakano \inst{1,2} \and
Celine Vens\inst{1,2}}

\authorrunning{A. Gharahighehi et al.}

\institute{KU Leuven, Campus KULAK, Department of Public Health and Primary Care, Kortrijk, Belgium \and Itec, imec research group at KU Leuven, Kortrijk, Belgium 
\email{\{alireza.gharahighehi,felipekenji.nakano,celine.vens\}@kuleuven.be}}

\maketitle              

\begin{abstract}
Recommender systems are information retrieval methods that predict user preferences to personalize services. These systems use the feedback and the ratings provided by users to model the behavior of users and to generate recommendations. Typically, the ratings are quite sparse, i.e., only a small fraction of items are rated by each user. To address this issue and enhance the performance, active learning strategies can be used to select the most informative items to be rated. This rating elicitation procedure enriches the interaction matrix with informative ratings and therefore assists the recommender system to better model the preferences of the users. In this paper, we evaluate various non-personalized and personalized rating elicitation strategies. We also propose a hybrid strategy that adaptively combines a non-personalized and a personalized strategy. Furthermore, we propose a new procedure to obtain \textit{free ratings} based on the side information of the items. We evaluate these ideas on the MovieLens dataset. The experiments reveal that our proposed hybrid strategy outperforms the strategies from the literature. We also propose the extent to which \textit{free ratings} are obtained, improving further the performance and also the user experience. 

\end{abstract}

\section{Introduction}
\label{sec1}
In the era of digitization, users use online platforms to buy products, read news, listen to music and watch movies. These online platforms typically provide a huge catalogue of items and services. In this situation, however, finding items of interest among many irrelevant ones is a hurdle for the users. Recommender systems (RSs) are intelligent retrieval methods that assist users in finding the relevant information. Generally, RSs can be divided into two main categories: content-based filtering (CB) and collaborative filtering (CF). 

As its name suggests, CB recommenders recommend items solely based on their content. That is, these methods will recommend items whose features match the user profile. Contrarily to CB, CF approaches utilize the interactions of other users to predict the preferences of the target users, consequently CF methods provide more precise~\cite{adomavicius2005toward} and less obvious~\cite{Lops2011} recommendations compared to CB. CF approaches face, however, the sparsity issue. 

The sparsity issue arises as users interact with a small fraction of the item catalog and therefore all the other possible interactions are missing. To address this issue, there are two main approaches: generating artificial logs via generative models, as suggested in~\cite{gharahighehi2021recommender}, or requesting real users to rate specific items, as performed by active learning (AL) strategies \cite{elahi2016survey}. Despite being vastly studied in other tasks, generating artificial data in RSs is still in its initial steps \cite{Slokom2018}. Thus, in this work, we explore further AL.

AL investigates strategies that select which item ratings from the users will most likely improve the performance of the RSs \cite{settles2010,elahi2016survey}. In a typical application, the most informative items will be sequentially selected until a stopping criterion is triggered, e.g., the performance of the RS has significantly improved.

AL strategies are roughly categorized into two groups: i) non-personalized and ii) personalized. Non-personalized strategies select the same items for all users. As a consequence, their recommendations are often redundant. On the other hand, personalized strategies request ratings from different items for each user, nevertheless they might not be applicable in some settings due to the aforementioned sparsity issue. 

As a possible solution to these drawbacks, we propose a new hybrid strategy. More specifically, our proposed AL strategy employs both non-personalized and personalized strategies in an adaptive fashion. Since data is very scarce in the first iterations, our strategy endorses the use of a non-personalized strategy. As more data becomes available in further iterations, the focus shifts to a personalized strategy.

Additionally, inspired by recent advances on AL for multi-label classification \cite{Wu2017,Wu2018}, we also propose an extra procedure where \textit{free ratings} can be obtained without extra costs. Based on the side information and item-embeddings, we identify the most similar items to the ones rated by the user, and rate them identically to the user ratings.
Its main objective consists of improving the performance without demanding more interaction from the user. Consequently, it also provides a better experience to the users since similar items will be automatically rated, dismissing the necessity of further rating. To the best of our knowledge, the concept of \textit{free ratings} has not been explored yet in the context of RSs.

Our experiments show that our proposed hybrid adaptive AL strategy outperforms the methods from the literature. Furthermore, we also show that the \textit{free ratings} obtained by our procedure are informative and capable of considerably improving the performance.

The contributions of this paper are summarized as follows:

\begin{enumerate}
    \item We propose a new hybrid AL strategy which combines the advantages of both personalized and non-personalized strategies;
    \item We provide an evaluation of several AL strategies using a CF method as the underlying recommender; 
    \item We also propose a new procedure that can automatically infer ratings from the user. 
\end{enumerate}


The remainder of this paper is organized as follows: In Section~\ref{sec2}, the related work on applications of AL in CF are reviewed. Next, in Section~\ref{sec3}, we present the background and describe different personalized and non-personalized AL strategies in CF. Then, in Section~\ref{sec4}, we propose a method to enhance the performance of rating elicitation procedure by dynamically combining two AL strategies and by using item side information. In Section~\ref{sec5}, we explain the experimental setup that we used to evaluate the AL strategies and the proposed approaches. Then, in Section~\ref{sec6}, several AL strategies in CF are evaluated using the MovieLens dataset. We also show in this section how the proposed approaches can enhance the performance of rating elicitation in CF. Finally, we conclude and outline some future directions in Section~\ref{sec7}.  

\section{Related work}
\label{sec2}

Sparsity is one of the major issues in CF. There are two main approaches in the literature to address sparsity: generating artificial user logs and rating elicitation~\cite{gharahighehi2021recommender}. Chonwiharnphan et al.~\cite{artificial_logs} proposed a generative model to generate artificial user logs. They used an auto-encoder to form item embeddings and then used them as input to a GAN-based to generate artificial user interactions. 

Rating elicitation is the natural mean of densifying the rating matrix between users and items. There are various AL strategies to densify the rating matrix in CF. Elahi et al.~\cite{elahi2016survey} provide a survey on AL strategies that have been used for rating elicitation in CF. They categorized these strategies w.r.t. personalization, i.e., whether the selected items are personalized or the same for all users and hybridization, i.e., whether the strategy is a single heuristic or a combination of multiple strategies. Chaaya et al.~\cite{chaaya2017evaluating} used this categorization and evaluated single heuristic non-personalized AL strategies in CF. They evaluated random, variance~\cite{rashid2002getting}, entropy~\cite{kohrs2001improving}, entropy0~\cite{rashid2008learning}, greedy extend~\cite{golbandi2010bootstrapping}, representative-based~\cite{liu2011wisdom}, popularity~\cite{kohrs2001improving} and co-coverage~\cite{golbandi2010bootstrapping} strategies and showed that entropy-based approaches have the best performance among these AL strategies.


In the context of multi-label classification, some works have proposed AL strategies to automatically infer labels from the unlabeled dataset \cite{Wu2017,Wu2018}. That is, given the labeled data provided by the oracle, and possibly side information, an AL algorithm would predict few additional labels from the unlabeled dataset. Consequently, more data is made available to the model, resulting in superior performance and reduced costs overall.     

For instance, Wu et al.\cite{Wu2017} proposed a three-component method where the prediction of the model itself, a K-Nearest neighbors classifier and a co-occurrence based label correlation component were combined to infer labels from the unlabeled dataset. Each component provides a single vote, and the final inferred label is decided based on the majoritarian vote. This method was further improved in \cite{Wu2018} where the co-occurrence label correlation criteria was replaced by a low-rank mapping matrix factorization method. In both cases, the most uncertain labels were selected to be inferred. To the best of our knowledge, this idea has not been investigated in RSs. 

In this paper we propose an adaptive hybrid strategy that hybridizes a non-personalized and a personalized AL strategies and uses some additional free ratings inferred from the unrated items. 
\section{Background}
\label{sec3}

In this paper, uppercase bold letters are representing matrices, lowercase bold letters are used for vectors, uppercase non-bold letters and lowercase non-bold letters represent sets and constants respectively. \(\textbf{M}_{i,j}\) represents the element in $i^{th}$ row and $j^{th}$ column of matrix \textbf{M}. 

The rating elicitation procedure can be online or offline~\cite{elahi2016survey}. As we do not have access to online users, we simulate the rating elicitation procedure via offline setting. Using the offline setting and notations proposed by Elahi et al.~\cite{elahi2016survey}, the dataset should be split into three disjoint matrices: $\textbf{K}$, $\textbf{X}$ and $\textbf{T}$. 

\begin{itemize}
    \item $\textbf{K}$: Matrix designed to simulate the initial, small and sparse dataset available in AL applications. $\textbf{K}$ is used to train the RS and its content is known by both the system and the user; 

    \item $\textbf{X}$: Matrix used for rating elicitation. It contains the ratings that can possibly be selected by the AL strategy. The ratings in this matrix are not known by the system but known by the user, i.e., the users can provide new ratings if requested;

    \item $\textbf{T}$: Matrix used only for evaluation; 
\end{itemize}

As aforementioned, the task of AL strategies is to find most relevant items to be rated by the user. This is performed in iterations. At the beginning of each iteration, the AL strategy selects the items to be rated from $\textbf{X}$. Such items are rated by the user, then included in $\textbf{K}$ and removed from $\textbf{X}$. The RS is rebuilt to include the freshly obtained data, and evaluated on $\textbf{T}$. This sequence of operations is repeated until a stopping criterion is triggered. $\textbf{T}$ remains unaltered.

As described by Elahi et al.~\cite{elahi2016survey}, AL strategies can be categorized in a taxonomy. From this taxonomy, we consider the following non-personalized strategies: \textit{variance}, \textit{entropy}, \textit{entropy0}, \textit{co-coverage}, \textit{popularity} and their hybridizations. The rating elicitation score of a candidate item based on these strategies can be calculated using the following equations (higher scores are preferred): 

\begin{equation}
Var(i) =\frac{1}{|U_{i}|}\sum_{u \in U_{i}}(r_{ui} - \overline{r_{i}})^{2}
\label{eq:variance}
\end{equation}

\begin{equation}
Ent(i)=-\sum_{r=1}^{5}\frac{|\{x| x \in U_{i}\cap \textbf{R}_{x,i}=r\}|}{|U_{i}|}log(\frac{|\{x| x \in U_{i} \cap \textbf{R}_{x,i}=r\}|}{|U_{i}|})\label{eq:entropy}
\end{equation}

\begin{equation}
Ent0(i)=- \sum_{r=0}^{5}\frac{|\{x|x \in U\cap \textbf{R}_{x,i}=r\}|}{|U|}log(\frac{|\{x| x \in U\cap \textbf{R}_{x,i}=r\}|}{|U|})
\label{eq:entropy0}
\end{equation}

\begin{equation}
Co-cov(i) =\sum_{j \in I \cap j\neq i}|\{x|x \in U_{i} \cap x \in U_{j}\}|\\
\label{eq:cov}
\end{equation}

\begin{equation}
Pop(i) =|U_{i}|
\label{eq:pop}
\end{equation}

\noindent where \(U_{i}\) is the set of users who have rated item \(i\), \(r_{ui}\) is the rating of user \(u\) to item \(i\), \(\overline{r_{i}}\) is the average rating of item \(i\), \(\textbf{R}\) is the rating matrix, \(U\) is the set of all users and \(I\) is the set of all items. 

There are also some hybrid non-personalized AL strategies that combine two of the mentioned strategies such as \textit{Pop\_entropy}, \textit{Pop\_variance} and \textit{HELF}.

\begin{equation}
Pop\_ent(i) = log(pop(i))\times Ent(i)
\label{eq:pop_ent}
\end{equation}

\begin{equation}
Pop\_var(i) = \sqrt{pop(i)}.Var(i)
\label{eq:pop_var}
\end{equation}

\begin{equation}
HELF(i)=\frac{2\times 
 log(|Pop(i)|)\times Ent(i)}{log(|U|)\times log(max(R)) \times (log(Pop(i))+\frac{Ent(i)}{log(max(R))})}
\label{eq:helf}
\end{equation}

All of these strategies are non-personalized and therefore the same set of items will be selected for all users. The other type of AL strategies is the personalized one where the choice of items to be rated is personalized for each user. We have evaluated \textit{MaxRating}, \textit{MinRating}, \textit{MinNorm}, \textit{IKNN}, \textit{Binary} and \textit{Non-myopic} strategies~\cite{elahi2016survey}. \textit{MaxRating} and \textit{MinRating} strategies use a predictor (for instance SVD) to predict the user ratings to the items and then select the items with higher predicted ratings and lower predicted ratings respectively. \textit{MinNorm} strategy selects the items whose Euclidean norms of latent features are the smallest. The \textit{IKNN} strategy is the common item-based CF method that forms neighborhood based on items. In the \textit{Binary} strategy, the rating matrix is transformed to a binary matrix, i.e., all the missing values are changed to zero and all the rest to one. This model predicts the likelihood that a user will rate an item, either a high rating or a low rating. Finally, \textit{Non-myopic} strategy adaptively combines \textit{MinRating} and \textit{MinNorm} strategies based on the number of iterations:    
\begin{equation}
Non-myopic(i)= w\times MinNorm\_R(i) + (1-w)\times MinRating\_R(i) 
\label{eq:A}
\end{equation}

\noindent where $MinNorm\_R(i)$ and $MinRating\_R(i)$ are the the ranks of item $i$ among the candidate items based on $MinNorm$ and $MinRating$ strategies respectively, and $w$ is a weight which adapts the effect of each strategy based on the number of proceeded rating elicitation iterations, i.e., $w = \frac{\#iter -1}{\#TotalIter}$.

\section{Our proposed method}
\label{sec4}
In this section, we explain our proposed method. First, in Section~\ref{method_hybrid}, we explain how we hybridize a non-personalized and a personalized AL strategy. Then, in Section~\ref{method_free}, we present the concept of free ratings where we use items features to add some additional (free) ratings to the model without asking users to rate more items.
\subsection{Adaptive Hybridization of Personalized and Non-personalized AL strategies}
\label{method_hybrid}

As mentioned in the previous sections, there are two main categories of AL strategies: personalized and non-personalized. The non-personalized strategies select the same items for all users while the personalized ones select different set of items for each user based on their previous interactions/elicited ratings. The choice between these two categories depends on knowledge of the model about user preferences. If the knowledge of the recommender about the target user is very limited, then the personalized AL strategies are not very effective as there is no means for personalization. Therefore, non-personalized strategies would be more logical in the very first steps of rating elicitation procedure. When the user provides more ratings, the system has a better view on the user preferences and therefore is able to perform personalized rating elicitation procedure. 

To utilize the potentials of both categories, based on our experiments (Section \ref{sec6}), we proposed an adaptive weighted hybrid approach based on the \textit{Pop\_entropy} strategy, which is a non-personalized AL approach, and the \textit{Binary} strategy, which is a personalized one. The hybridization weight in the proposed approach is adaptive based on the stage of rating elicitation procedure. In the very early stages, the hybrid approach depends more on the \textit{Pop\_entropy} strategy and when the system gains more knowledge about user preferences it assigns more weight to the \textit{Binary} approach. The rating elicitation score of a candidate item for a user in the proposed hybrid approach is calculated based on the following equation:

\begin{equation}
Hybrid(u,i)=w\times Pop\_ent\_R(i) + (1-w)\times Binary\_R(u,i)
\label{eq:hybrid}
\end{equation}

\noindent where \(Pop\_ent\_R(i)\) is the ranking of item \textit{i} based on the \textit{Pop\_entropy} strategy (Eq.\ref{eq:pop_ent}), \(Binary\_R(u,i)\) is the ranking of the item \textit{i} for user \textit{u} based on the \textit{Binary} strategy and \textit{w} is the dynamic weight. The weight should shift from the non-personalized approach to the personalized one when some few ratings have been elicited from the user. To model this, we used an exponential hybridization weight:

\begin{equation}
w=exp^{-\alpha \times \# iter }
\label{eq:w_hybrid}
\end{equation}

\noindent where $\#iter$ represents the iteration round and \(\alpha\) is a hyperparameter. Higher values of $\alpha$ shift the weight from \textit{Pop\_entropy} to \textit{Binary} faster. 

\subsection{Rating elicitation with \textit{free ratings}}
\label{method_free}

Requesting multiple ratings from the user can backfire. In some cases, it can lead to disengagement and an overall uncomfortable experience. To address this issue, and possibly enhance the performance, we hypothesize that more information can be obtained from the ratings of the user. 

In several applications, items can be described using features that represent their content. For instance, in a movie streaming platform each movie can be represented by its genre and artists. Complementary to these explicit features, model-based CF methods learn item latent features, which we call item embeddings, from the interaction matrix. We propose to employ explicit features and embeddings of items to obtain \textit{free ratings}.

More specifically, we assume that similar items will be similarly rated by the same user. That is, for each item in a set of selected items provided by an AL strategy, we find their most similar items, and provide them the same rate. Afterwards, these automatically inferred ratings are incorporated to the model, obtaining, thus, \textit{free ratings} at each iteration. In order to find similar items, we have employed the cosine similarity, as it is often used in RSs context. Given that item $i$ is selected by the AL strategy for rating elicitation, $\textbf{f}_{i}$ is the explicit feature vector and $\textbf{e}_{i}$ is the learned embedding of item $i$, the free rating related to this item is:

\begin{equation}
free(i) = \argmax_{j \in I \cap j\neq i} cosine\_sim (\textbf{v}_{i},\textbf{v}_{j})
\label{eq:free_r}
\end{equation}

\noindent where $\textbf{v}_{i}$ is the concatenation of $\textbf{f}_{i}$ and $\textbf{e}_{i}$. The Pseudocode of the proposed adaptive hybrid rating elicitation strategy with the \textit{free ratings} procedure is illustrated in Algorithm 1. Using the hybridization strategy, the proposed approach can take the advantages of both personalized and non-personalized AL strategies. Moreover, exploiting the \textit{free ratings} empowers the proposed AL strategy to gather enriched information when eliciting real ratings from users.

\begin{algorithm}
    \caption{Adaptive hybrid AL strategy with free ratings}
    \SetKwData{Left}{left}\SetKwData{This}{this}\SetKwData{Up}{up}
    \SetKwFunction{Union}{Union}\SetKwFunction{FindCompress}{FindCompress}
    \SetKwInOut{Input}{input}\SetKwInOut{Output}{output}

    \Input{dataset, $\#TotalIter, n, RS$  \tcp{The base RS can be any recommender, here it is SVD. $n$ is the number of neighbors.}}
    \Output{items to be rated by each user, MAE per round}
    \BlankLine
    \emph{$K, X, T \gets$split the dataset \tcp{$K$ for training, $X$ for rating elicitation and $T$ for evaluation.}}
    \emph{$model \gets$train the RS on $K$}\;
    \emph{$metric  \gets$evaluate $model$ on $T$}\;
    \emph{$j\gets$ 1}\;
    \While{$j<= \#TotalIter$}{
         $Pop\_ent\_recom\gets Pop\_ent(X)$ (Eq.\ref{eq:pop_ent})\tcp{Selected items by \textit{Pop\_entropy} strategy are the same for all users.}\
        \For{$u \in U$ }{\label{forins}
             $ S_{u} \gets$ $hybrid(u,Pop\_ent\_recom)$ (Eq.\ref{eq:hybrid}) \;
             $ R_{u} \gets top@10(Sorted(S_{u}))$ \;
             $ K_{u} \gets K_{u} \bigcup R_{u}$ \;
             $ X_{u} \gets X_{u} / R_{u}$\;
            \For{$i \in R_{u}$}{\label{forins}
                 $ f\_rating \gets$ $free(i)$ (Eq.~\ref{eq:free_r})\ 
                
                \For{$f \in f\_rating$}{\label{forins}
                     $ K_{uf} \gets$ $K_{ui}$\;
              
                }  
                }
            }
            
             $model \gets$train the RS on K\;
             $metric \gets$evaluate model on T\;
     $j \gets j +1$\;
    }
           \label{alg:simulation}
\end{algorithm}

\section{Experimental Setup}
\label{sec5}
In this section, we present details about the dataset employed, base RS, comparison methods, and evaluation measures
\subsection{Dataset}
To evaluate various AL strategies and the proposed method, we used the MovieLens 1M~\cite{harper2015movielens} dataset. This dataset contains around 1 million ratings (from 1 to 5) of users to the movies. The movie features, including movie genres and artists, are used to select free ratings. This dataset contains the ratings of 6,0640 users on 3,706 movies. Movies belong to 18 genres and contain 24,853 unique actors. To evaluate the evolution of rating elicitation procedure, we have filtered the dataset to include only users and items with at least 100 ratings. 

We have selected Movielens dataset as it is a standard dataset in the RS community and it comes with rich item side information. As the idea of adding free ratings relies on rich features, this publicly available dataset is ideal for benchmarking the proposed method.

As mentioned in Section~\ref{sec3}, we split the dataset to three disjoint matrices, namely $\textbf{K}$, $\textbf{X}$ and $\textbf{T}$. In this paper, we start with a very sparse matrix for $\textbf{K}$ which only contains one rating per user. We keep 30 ratings per user in matrix $\textbf{T}$\footnote{Each user can have ratings of different items in the test set.} for evaluation and the remaining ratings are used for applying AL strategies in matrix $\textbf{X}$. 

\subsection{Base recommender system}
To evaluate the quality of the rating elicitation procedure, a CF method is needed to incrementally be trained on the selected ratings by the AL strategy. In this paper, we use \textit{SVD} (Singular Value Decomposition)~\cite{salakhutdinov2008bayesian} as the CF method. Naturally, other CF methods could also have been applied, nonetheless our focus relies on investigating the effect of AL strategies on the performance of the CF method. For a complete review on CF methods, we refer the reader to~\cite{shi2014collaborative}.

\subsection{Comparison methods}
We compared the methods over 25 iterations. At each iteration, 10 items per user are requested to be rated. 

\subsubsection{AL strategies}
We have evaluated the following AL strategies:
\begin{itemize}
    \item \textbf{Non-personalized:} variance, entropy, entropy0, popularity, co-rating, pop\_variance, pop\_entropy and Helf;
    \item \textbf{Personalized}: min\_rating, max\_rating, min\_norm, non-myopic, IKNN\footnote{We used 40 for k.} and binary\footnote{We used 291 latent factors,1501 iterations, 0.01834 for learning rate and 0.01467 for regularization.}; 

    \item \textbf{Adaptive hybrid:} This is the proposed method that adaptively hybridizes binary and pop\_entropy strategies. We have set $\alpha$ to 2;
\end{itemize}

\subsubsection{Free ratings}
As mentioned before, this is the first study to address \textit{free ratings} in the context of RSs. Hence, to the best of our knowledge, the literature does not contain any comparison method.

We have evaluated two variants of our procedure. Their difference relies on the representations used to find the most similar items:

\begin{itemize}
    \item Only features: only the features available as side information;
    \item Features + embeddings: the side information features concatenated with the item-embeddings obtained via the decomposition performed by SVD;
\end{itemize}

A high number of \textit{free ratings} could contaminate the dataset. Since the \textit{free ratings} are inferred and not actually rated by the user, it is necessary to balance the trade-off between free information and possible noise.

Hence, the number of \textit{free ratings} per rated item is equal to 1, meaning that only the most similar item of each item user rated item is included.

\subsection{Evaluation measure}
To evaluate the performance of rating elicitation procedure, we use the MAE (Mean Absolute Error) measure, since it is vastly employed in the literature~\cite{elahi2016survey,chaaya2017evaluating}.

\begin{equation}
MAE=\frac{1}{|\textbf{T}|}\sum_{r_{ui}\in \textbf{T}}|r_{ui} - \hat{r}_{ui}|
\label{eq:MAE}
\end{equation}

 In Equation (\ref{eq:MAE}) $r_{ui}$ is the real rating of item $i$ given by user $u$, $\hat{r}_{ui}$ is the predicted rating using the CF method and $\textbf{T}$ is the evaluation matrix.

\section{Results and Discussion}
\label{sec6}

To better interpret our experiments, we present the results in separate subsections.

Initially, we compare all personalized and non-personalized methods among themselves. Next, to validate our proposal, we compare the best methods of the two categories against our proposed method. Finally, we perform experiments to validate our hypothesis about \textit{free ratings.}

All figures presented in this section are best viewed in colors.

\subsection{Personalized vs non-Personalized}

\begin{figure}[!htb]
\centering
\includegraphics[width=0.7\linewidth]{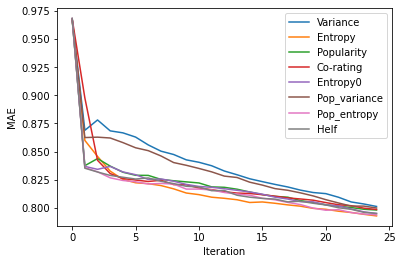}
  \caption{Evaluation of non-personalized AL strategies}\label{fig:non}
\end{figure}

The comparison between various non-personalized AL strategies is illustrated in Figure~\ref{fig:non}. As shown in this figure, most of the methods performed in a very overlapping fashion. Among these strategies, \textit{Pop\_entropy} has the best performance in the initial steps and \textit{Entropy} is superior in later iterations (from the seventh iteration). 

Differently from that, the strategy \textit{variance} and \textit{pop\_variance} underperformed. In very few iterations, it is noticeable that the selected ratings are not as relevant, making them not competitive with the other strategies.

\begin{figure}[!htb]
\centering
  \includegraphics[width=0.7\linewidth]{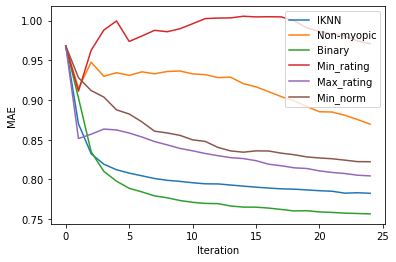}
  \caption{Evaluation of personalized AL strategies}\label{fig:pers}
\end{figure}

When it comes to the personalized AL strategies, \textit{Binary} has the best performance according to Figure~\ref{fig:pers}. Despite being slightly overpowered by \textit{IKNN} and \textit{Max\_rating} in the first iterations, Binary is mostly associated to the superior performance.

As expected, \textit{Min\_rating} and its hybridization (\textit{Non-myopic}) have inferior performance. This strategy is based on an exploratory approach where the lowest rated items are selected. The rationale behind that relies on the hypothesis that low rated items contain novel and relevant information. But, as shown in Figure~\ref{fig:pers}, the results does not support this assumption.

In fact, such undesired behaviors are often associated to random and non-informative ratings. Normally, an exponential improvement is perceived in the first iterations, as yielded by \textit{Binary} and \textit{IKNN}. However, a linear and even degrading performance is observed in \textit{Min\_rating}, \textit{Non-myopic}, \textit{Max\_rating} and \textit{Min\_norm}. Such findings are rather surprising since such strategies are employed as comparisons.

Hence, \textit{Binary} and \textit{Pop\_entropy} perform the best among the personalized and non-personalized AL strategy respectively.

\subsection{Adaptive hybrid strategy}
\begin{figure}[!htb]
\centering
  \includegraphics[width=0.7\linewidth]{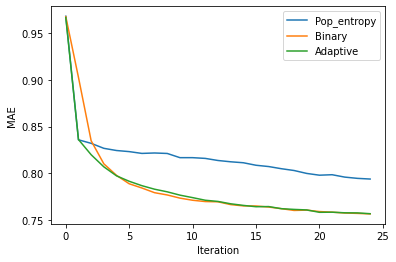}
  \caption{Evaluation of adaptive hybrid AL strategy}\label{fig:adaptive}
\end{figure}

At Figure \ref{fig:adaptive}, we present a comparison between our proposed method (\textit{adaptive}) and the two best methods from the personalized (\textit{Binary}) and non-personalized (\textit{Pop\_entropy}) approaches.

As expected, the Binary method performed poorly in the first iterations due to the lack of data for personalization, nonetheless it managed to replicate the performance of our method after some iterations.

Likewise, the non-personalized method, \textit{Pop\_entropy}, outperformed its personalized counterpart in the first iterations. Its performance, however, rapidly deteriorates thereafter.    

As opposed to that, our method yielded superior performance by utilizing the advantages of its both components. In the beginning, its non-personalized one is capable of overcoming the lack of data, whereas its personalized part guarantees that the performance does not stagnate.

Despite its simplicity, a shifting weight is capable of combining the advantages of both personalized and non-personalized approaches.

\subsection{\textit{Free ratings}}

\begin{figure}[!htb]
\centering
  \includegraphics[width=0.7\linewidth]{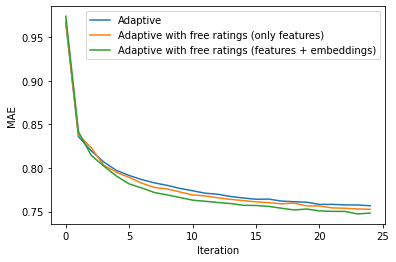}
  \caption{Adding free ratings to the adaptive hybrid approach}\label{fig:free}
\end{figure}

As shown in Figure \ref{fig:free}, incorporating \textit{free ratings} is associated to superior performance. In a very few iterations, the baseline, \textit{adaptive} in this case, is rapidly surpassed  by the other two methods.

A more prominent difference is noticed from iteration 5 onwards where using the features and the embeddings concatenated (green) is visibly superior, followed by its simpler counterpart which only employs the features (orange). This behavior is firmly maintained throughout all iterations.

Despite being a straightforward solution, using only the features is not associated to the best performance. We recommend thus concatenating them with the embeddings.  This could be interpreted as an indication that our proposed method is applicable in other datasets as well, since item embeddings are obtainable in most of the contexts, differently from side information.    

These results confirm our hypothesis. \textit{Free ratings} are obtainable in the context of RSs. They are not only an extra asset to improve the performance, but they may also be used to refine the user experience, since requesting ratings can become monotonous.

\section{Conclusion and Future work}
\label{sec7}
In this paper, multiple non-personalized and personalized AL strategies in CF were assessed. It has been showed that \textit{Pop\_entropy} and \textit{Binary} are the most effective non-personalized and personalized AL strategies respectively in the evaluated dataset. Non-personalized strategies are more effective when the RS has very limited knowledge about a user, whereas personalized strategies are able to select more relevant items when the system has deeper knowledge about that user. 

We also proposed an adaptive hybrid AL strategy that combines \textit{Pop\_entropy} and \textit{Binary} strategies to utilize their merits. In the initial steps of rating elicitation procedure, it puts more focus on \textit{Pop\_entropy} and, in the later steps, \textit{Binary} strategy plays the main role in selecting candidate items. 

We have also proposed a procedure to automatically infer ratings, referred to as \textit{free ratings}, based on the item features and embeddings. Our results show that such \textit{free ratings} are capable of improving the performance of the model without requesting further interacting from user, improving, thus, his/her experience.

We outline the following directions for future work:

\begin{itemize}
    \item Further validation: As a proof of concept, we have restricted our experiments to the MovieLens dataset due to the presence of side information. Future work should consider datasets from other applications. 
    \item Further adaptability for our hybrid strategy: A future version of our proposed hybrid adaptive strategy could automatically identify which personalized and non-personalized strategy should be included. This could be performed by means of an validation subset.
    \item Adaptive \textit{free ratings:} We have estimated the value of the \textit{free ratings} by attributing them the same value as the real ratings. These additional ratings are not real ratings and the RS is less confident about their real values. Therefore, one could consider using a smaller weight instead of the same value.
    \item User adaptive strategy: The weight responsible to shift the focus from non-personalized to personalized is determined by the current iteration. The number of rated items, however, might vary from user to user. A future version could employ the number of rated items by the user in question. 
    \item \textit{Free ratings} from the training dataset: It is also possible \textit{free ratings} from the whole rated dataset, instead of only using the freshly obtained ratings. This change could lead to further improvement in the performance.
    
    \item Diversity: A diverse list of items is often associated to better performance. In some domains such as news~\cite{gharahighehi2021diversification,gharahighehi2020making} and music~\cite{gharahighehi2021personalizing} recommendations, diversity is an important factor that should be considered in the model.  
    \item Implicit feedback: We have relied on explicit feedback from users. However, in reality, usually users passively express their preferences by using and consuming the services in online platforms~\cite{gharahighehi2019extended}. Therefore the feedback elicitation should be adapted to capture this implicit feedback from users.  
     
    \item Multiple stakeholders: In some contexts, such as job finding or news aggregator websites~\cite{gharahighehi2020multi,gharahighehi2021fair}, there are multiple stakeholders in the system and therefore in the rating elicitation procedure the preferences of different stakeholders should be considered.
\end{itemize}

\subsubsection*{Acknowledgements} This work was executed within the Immosite.com project, an innovation project co-funded by Flanders Innovation \& Entrepreneurship (project nr. HBC.2020.2674) and with involvement from industrial partners Immosite and g-company. The authors also acknowledge support from the Flemish Government (AI Research Program).

\bibliographystyle{template/bibtex/spmpsci}
\bibliography{template/bib} 

\end{document}